\documentclass[a4paper]{article}
\pdfoutput=1
\usepackage{jheppub}
\usepackage{float} 

\usepackage[utf8]{inputenc}

\usepackage[table ]{ xcolor}

\usepackage{multirow}

\usepackage{amsmath}

\usepackage{tikz}
\usetikzlibrary{shapes,arrows}
\tikzstyle{startstop} = [rectangle,rounded corners, minimum width=3cm,minimum height=1cm,text centered, draw=black,fill=red!30]
\tikzstyle{io} = [trapezium, trapezium left angle = 70,trapezium right angle=110,minimum width=3cm,minimum height=1cm,text centered,draw=black,fill=blue!30]
\tikzstyle{process} = [rectangle,minimum width=3cm,minimum height=1cm,text centered,text width =3cm,draw=black,fill=orange!30]
\tikzstyle{decision} = [diamond,minimum width=3cm,minimum height=1cm,shape aspect=3,inner sep = 0.4pt,text centered,draw=black,fill=green!30]
\tikzstyle{arrow} = [thick,->,>=stealth]
\tikzstyle{shadow}=[preaction={fill=black,opacity=.5,transform canvas={xshift=0.5mm,yshift=-0.5mm},shading=radial,shading angle=20},fill=red]

\tikzstyle{ellipse}=[draw, rectangle, minimum width=2.8em, rounded corners=6pt,line width=0.5pt]
\tikzstyle{pxsbx}=[trapezium, trapezium left angle=75, trapezium right angle=105, minimum width=3em, text centered, draw = black, fill=white,line width=0.5pt] 
\tikzstyle{lingxing}=[draw,diamond,shape aspect=3,inner sep = 0.4pt,thick,font=\itshape,line width=0.5pt]

\usepackage{amssymb}
\usepackage{graphicx,color}
\usepackage{appendix}
\usepackage{amsmath,amsthm}

\usepackage{mathrsfs}

\usepackage{bm}

\def\beq{\begin{equation}}
\def\eeq{\end{equation}}
\newcommand{\bea}{\begin{eqnarray}}
\newcommand{\eea}{\end{eqnarray}}
\def\bi{\begin{itemize}}
\def\ei{\end{itemize}}
\def\ba{\begin{array}}
\def\ea{\end{array}}
\def\bfig{\begin{figure}}
\def\efig{\end{figure}}

\def\be{\begin{eqnarray}}
\def\ee{\end{eqnarray}}

\renewcommand{\o}{\omega}

\newcommand{\dd}{\mathrm{d}}

\title{Quantum Oppenheimer-Snyder black hole evaporation and its fate}

\author[1]{\ Hongwei Tan}
\author[1]{Kui Xiao}

\author[2,3]{\ Rong-zhen Guo}  

\author[1]{\ Shoucheng Wang}

\affiliation[1]{ School of Science, Hunan Institute of Technology, Hengyang 421002, China}

\affiliation[2]{School of Fundamental Physics and Mathematical Sciences Hangzhou Institute for Advanced Study, UCAS,
Hangzhou 310024, China}

\affiliation[3]{School of Physical Sciences, University of Chinese Academy of Sciences, No. 19A Yuquan Road, Beijing
100049, China}

\emailAdd{honweitan@hnit.edu.cn}
\emailAdd{Corresponding author:xiaokui@hnit.edu.cn}
\emailAdd{Corresponding author:guorongzhen@ucas.ac.cn}
\emailAdd{scwang@hnit.edu.cn}

\abstract{In this paper, we investigate the evaporation of the quantum Oppenheimer-Snyder black hole. Within a semiclassical framework, we compute the energy emission of Hawking radiation by introducing a massless scalar field as a test field, considering both minimally and non-minimally coupled cases. For the minimally coupled case, we find that loop quantum gravity effects become crucial at the late stage of the evaporation process, causing the emission rate to slow down and eventually terminate, leading to the formation of a black hole remnant. A quasi-normal mode analysis indicates the stability of this remnant.
For the non-minimally coupled case, we show that the fate of the black hole strongly depends on the value of the coupling constant $\xi$.
Focusing on the cases $\xi=\pm1$, we find that for $\xi=1$, the energy emission rate accelerates at late times and no remnant is formed, whereas for $\xi=-1$, the emission rate slows down and eventually terminates, resulting in a stable black hole remnant, as supported by the corresponding quasi-normal mode analysis.
}

\keywords{}

\begin{document}

\maketitle

\section{Introduction}
Black holes (BHs) are predicted by  General Relativity (GR) and play a central role in modern physics (see e.g., \cite{wald2010general, poisson2004relativist,frolov2011introduction,afshordi2024black,chandrasekhar1985mathematical}).
With the development of the observational technologies, the existence of the BHs has been confirmed by various observational signals, ranging from electromagnetic radiation to gravitational waves \cite{event2019first, akiyama2019first, akiyama2019first-2,akiyama2019first-3,akiyama2019first-4,akiyama2019first-5,LIGOScientific:2016aoc,LIGOScientific:2016sjg,LIGOScientific:2016dsl,LIGOScientific:2017bnn,abbott2016observation}.

In the 1970s, Bekenstein and Hawking proposed BH thermodynamics \cite{bekenstein1973black,hawking1975particle}. 
In this framework, a BH is treated as a thermodynamic system that emits particles, known as the Hawking radiation.
The entropy of a BH in Bekenstein-Hawking proposal is proportional to the BH area 
\begin{equation}
    S=\frac{\mathscr{A}}{4}.\footnote{In this paper, we choose the natural unit system $C=\hbar=G=k_{\text{B}}=1$, with $k_{\text{B}}$ is the Boltzmann constant.}
\end{equation}
The discovery of the BH thermodynamics opened a new window to understand the BH physics and provided profound insights into the nature of quantum gravity.
However, in the original BH thermodynamics proposal, the Hawking radiation is purely black body radiation, which is a mixed state.
Thus, Hawking radiation leads to a non-unitary evolution of the BH, resulting in the well-known BH information paradox (see e.g., \cite{mathur2009information,raju2022lessons,witten2025introduction}).
In recent decades, several approaches have been proposed attempting to resolve the information paradox \cite{susskind2012singularities,almheiri2021entropy,umemoto2018entanglement,engelhardt2022canonical,page1993information,page2013time,parikh2000hawking,parikh2004secret,zhang2008black,zhang2009black,zhang2025information}.

Loop quantum gravity (LQG), a background independent and non-perturbative approach to quantum gravity, potentially provides a resolution to the BH information paradox \cite{thiemann2008modern, rovelli2004quantum,han2007fundamental,thiemann2003lectures,ashtekar2004background,giesel2012classical,rovelli2011zakopane,perez2013spin}.
For instance, the black-to-white hole transition suggests a long-life remnant that can store the information of the collapsing matter, thereby resolving the information paradox \cite{haggard2015quantum,han2023geometry2,bianchi2018white}.
More recently, the evaporation process of the BH with holonomy corrections from LQG has been investigated \cite{belfaqih2025hawking}.
In that work, the authors demonstrated that the LQG corrections play a crucial role in the late stages of the BH evaporation by suppressing the Hawking radiation emission rate, thereby suggesting the formation of a BH remnant.

Another important effective BH model in the framework of LQG is the quantum Oppenheimer-Snyder (q-OS) BH \cite{lewandowski2023quantum}, which is a quantum extension of the classical Oppenheimer-Snyder model-the first model describing self-gravitational collapse, see \cite{oppenheimer1939continued}.
In the q-OS model, the spacetime is divided into an interior and an exterior region.
The interior region is filled with collapsing matter, modeled as a homogeneous and isotropic dust field,
and is described by the typical Friedmann-Robertson-Walker (FRW) metric $\mathrm{d}s^2=-\mathrm{d} \tau^2+a(\tau)^2\left(\mathrm{~d} \tilde{r}^2+\tilde{r}^2 \mathrm{d} \Omega^2\right)$.
The scalar factor $a(\tau)$ satisfies the LQG-modified Friedmann equation, which is known as the Ashtekar-Pawlowski-Singh (APS) model \cite{ashtekar2006quantum}, $H^2:=\left(\frac{\dot{a}(\tau)}{a(\tau)}\right)^2=\frac{8 \pi }{3} \rho\left(1-\frac{\rho}{\rho_c}\right)$, where $H$ is the Hubble constant
and $\rho_c$ is called the critical density.
In the semi-classical limit, $\rho_c\to\infty$, the equation of motion (EOM) for $a(\tau)$ reduces to the standard  Friedmann equation $H^2:=\left(\frac{\dot{a}(\tau)}{a(\tau)}\right)^2=\frac{8 \pi }{3} \rho$.
The metric of the exterior spacetime region is given by applying the  Israel junction conditions \cite{israel1966singular,shi2024higher}
\begin{equation}\label{eq:quan_OS}
    \mathrm{d} s_{\text{q-OS}}^2
    =-F(r)\mathrm{d}t^2
    +F(r)^{-1}\mathrm{d}r^2
    +r^2\dd \theta^2
    +r^2\sin^2\theta\dd\varphi^2.
\end{equation}
Here, $(t,r)$ are the usual time and radial coordinates of the exterior spacetime region, and $F=1-\frac{2M}{r}+\frac{\alpha M^2}{r^4}$.
The q-OS model  thus provides an effective description within LQG, with the quantum correction term $\frac{\alpha M^2}{r^4}$ arising from the LQG effects.
In the semi-classical limit $\alpha\to0$, the q-OS model \eqref{eq:quan_OS} reduces smoothly to the classical Schwarzschild BH. 

The quantum correction term $\frac{\alpha M^2}{r^4}$ leads to a variety of   interesting physical implications (see e.g., \cite{yang2023shadow,stashko2024quasinormal,zhang2023black,gong2024quasinormal,yang2024gravitational,liu2024gravitational,zi2024eccentric}).
In particular, in our previous works \cite{tan2025black,tan2025massive}, we investigated the tunneling processes of the q-OS BH with the Parikh-Wilczek approach \cite{parikh2000hawking, parikh2004secret}.
As a result, we found that the entropy of the q-OS BH is
\begin{equation}\label{eq:BH_entropy}
S_{\text{q-OS}}=S_{\text{Sch}}+k\log\left(\mathscr{A}_{\text{Sch}}\right).
\end{equation}
Here, $S_{\text{Sch}}$ denotes the entropy of the classical Schwarzschild BH, $\mathscr{A}_{\text{Sch}}$ is the area of the classical Schwarzschild BH, and $k$ is a constant that depends on the parameter $\alpha$.
Compared with the classical BH entropy, Eq. \eqref{eq:BH_entropy} exhibits a logarithmic correction in the black hole area for the q-OS black hole.
This result is consistent with the universal logarithmic corrections to black hole entropy that arise when quantum gravity effects are taken into account \cite{ghosh2005log,kaul2000logarithmic,domagala2004black,meissner2004black,chatterjee2004universal,medved2004comment,lin2024effective, shi2024higher,banerjee2008quantum,banerjee2008quantum2,banerjee2009quantum,majhi2009fermion,majhi2010hawking}).

In this work, we present another interesting physical implication arising from the quantum correction $\frac{\alpha M^2}{r^4}$.
By introducing a massless scalar field as a test field, we investigate the evaporation process of the q-OS BH.
Both the minimally coupled cases and the non-minimally coupled cases are considered.
For the minimally coupled cases, we find that the energy emission rate gradually slows down and eventually terminates,
indicating the formation of a  remnant of the q-OS BH evaporation.
Furthermore, by applying the higher-order WKB manner introduced in \cite{iyer1987black}, we compute the quasi-normal modes (QNMs), which indicates the stability of the BH remnant.
For the non-minimally coupled cases, we consider a massless scalar field satisfying the EOM $\nabla^a\nabla_a\phi-\xi g(r)R\phi=0$, where $R$ is the Ricci scalar of the spacetime and $\xi$ denotes the coupling constant.
$g(r)$ is a function of the radial coordinate $r$, which satisfies the following requirements: i) $\lim_{r\to r_{\text{H}}}g(r)=\lim_{r\to r_{\text{H}}}F(r)^{-1}$, with $r_{\text{H}}$ being the location of the event horizon; ii) $g(r)$ is finite when $r$ is significantly larger than $r_{\text{H}}$; iii) $\lim_{r\to\infty}g(r)$=1.
We focus on the cases of $\xi=\pm1$. 
For $\xi=1$, we find that the energy emission rate of the Hawking radiation does not slow down and no remnant is formed.
In contrast, for $\xi=-1$, the energy emission rate again slows down and eventually terminates, leading to the formation of a BH remnant, similar to the minimally coupled case.
The corresponding QNMs also imply the stability of the BH remnant.

Our results imply that the BH remnants formed during the evaporation process due to LQG corrections, consistent with previous results in Ref. \cite{belfaqih2025hawking}.
These results further indicate that the LQG effects potentially resolve the BH information paradox, as the information of the collapsing matter is expected to be stored in the BH remnant.

This paper is organized as follows: In Sec. \ref{sec:min}, we compute the Hawking radiation of the q-OS BH by considering a massless minimally coupled  scalar test field.
 In Sec. \ref{sec:non-min}, we analyze the Hawking radiation for a massless non-minimally coupled scalar test field.
 Conclusions and discussions are presented in Sec. \ref{sec:con}.

\section{Black hole evaporation with minimally coupled scalar field}\label{sec:min}
\subsection{The equation of motion of the minimally coupled scalar field}
In this section, we analyze the BH evaporation of the q-OS BH.
We first consider the test fields as a massless minimally coupled scalar field $\phi$, whose EOM reads
\begin{equation}
    g^{ab}\nabla_a\nabla_b\phi=0.
\end{equation}
Here, $g_{ab}$ is the q-OS metric, and $\nabla_a$ is the covariant derivative operator compatible with $g_{ab}$, satisfying $\nabla_ag_{bc}=0$.
According to the spacetime line element \eqref{eq:quan_OS}, this EOM can be expressed as 
\begin{equation}\label{eq:EOM_Non}
    -(
    1
    -\frac{2M}{r}
    +\frac{\alpha M^2}{r^4}
)^{-1}\ddot{\phi}
+\frac{2}{r}
(
    1
    -\frac{M}{r}
    -\frac{\alpha M^2}{r^4}
)\phi'
+(1-\frac{2M}{r}+\frac{\alpha M^2}{r^4})\phi''
+\frac{1}{r^2}\Delta^\theta\phi=0.
\end{equation}
Here, $\dot{\phi}$ and $\phi'$ represent the derivatives of  $\phi$ respect to $t$ and $r$, respectively.
$\Delta^\theta$ is the Laplacian on a unit 2-sphere, which reads 
\begin{equation}
    \Delta^\theta 
    :=\frac{1}{\sin\theta}\partial_\theta[\sin\theta\partial_\theta]
    +\frac{\partial^2_\varphi}{\sin^2\theta}.
\end{equation}
The test field $\phi$ can be expanded by the real spherical harmonics as 
\begin{equation}
    \phi(t,r,\theta,\phi)
    =\sum_{l,m}\tilde{\phi}_{lm}(t,r)Y_{lm}(\theta,\phi).
\end{equation}
Substituting this expansion into the \eqref{eq:EOM_Non}, we obtain
\begin{equation}\label{eq:EOM_non_mode}
    -(
    1
    -\frac{2M}{r}
    +\frac{\alpha M^2}{r^4}
)^{-1}\ddot{\tilde\phi}_{lm}
+\frac{2}{r}
(
    1
    -\frac{M}{r}
    -\frac{\alpha M^2}{r^4}
)\tilde\phi_{lm}'
+(1-\frac{2M}{r}+\frac{\alpha M^2}{r^4})\tilde\phi_{lm}''
-\frac{l(l+1)}{r^2}\tilde\phi_{lm}=0.
\end{equation} 
Here, we have used $ \Delta^\theta Y_{lm}=-l(l+1)Y_{lm}$.
To further simplify the \eqref{eq:EOM_non_mode}, we introduce the field redefinition 
\begin{equation}
    \tilde{\phi}_{lm}(t,r)
    =\frac{u_{lm}(t,r)}{r}.
\end{equation}
We then perform a Fourier transformation with respect to the time coordinate,
\begin{equation}
    u_{lm}(t,r)=\int_{-\infty}^\infty\tilde{u}_{lm}(t,\o)\frac{\dd\o}{2\pi}e^{-i\o t}.
\end{equation}
Next, we define the tortoise coordinate $r_*$ as
\begin{equation}\label{eq:def:tortoise}
    \dd r_*=\frac{\dd r}{F(r)}.
\end{equation}
In terms of the tortoise coordinate, the radial equation reduces to a Schrödinger-like form,
\begin{equation}\label{eq:EOM_non_tortoise}
\frac{\partial^2}{\partial r^2_*}u_{lm}
+(
    \omega^2
-V_l(r)
)u_{lm}
=0,
\end{equation}
with 
\begin{equation}\label{eq:def:eff_potential}
    V_l(r)=-(
    1
    -\frac{2M}{r}
    +\frac{\alpha M^2}{r^4}
)(\frac{4\alpha M^2}{r^6}
-\frac{l(l+1)}{r^2}
-\frac{2M}{r^3}).
\end{equation}
\subsection{The greybody factor and the black hole evaporation}
\begin{figure}[H] 
	\centering 
	\includegraphics[width=0.4\textwidth]{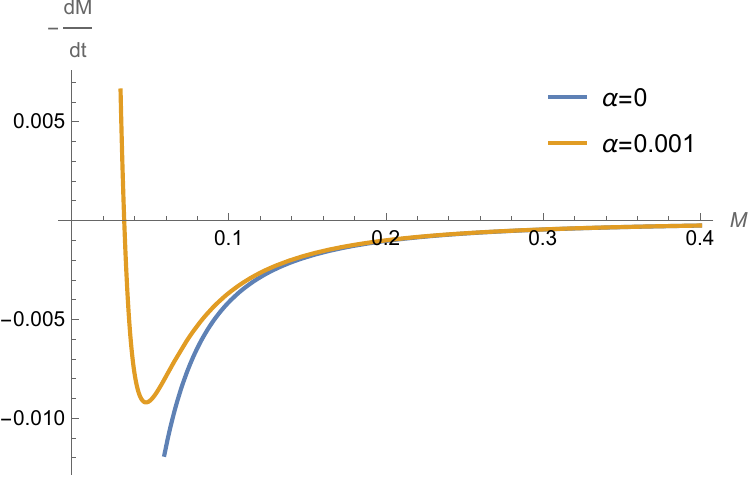} 
	\caption{Comparison of the Hawking radiation emission rate between the classical Schwarzschild black hole and the q-OS black hole with a minimally coupled scalar field.} 
	\label{fig:minimal_rate} 
	\end{figure}

In this paper, we neglect the back reaction of the emitted particles.
Furthermore, we follow the low frequency limit in \cite{harmark2010greybody}, which reads
\begin{equation}
    \o\ll T_{\text{H}},\,
    \o r_{\text{H}}\ll 1,
\end{equation}
where $T_{\text{H}}$ denotes the Hawking temperature, and
$r_H$ is the radius of the event horizon, given by \cite{tan2025black} 
\begin{equation}\label{eq:hor_rad}
    r_{\text{H}}\approx 2M-\frac{\alpha}{8M}.
\end{equation}
Then, the Hawking radiation spectrum is given by 
\begin{equation}\label{eq:eq:haw_spe}
    <N_\o>=\frac{T_l(\o)}{\exp(\o /T_{\text{H}})-1}.
\end{equation}
Here, $T_l(\o)$ is the greybody factor. 
The Hawking radiation is computed as \cite{shi2024higher}
\begin{equation}
    T_{\text{H}}=\left.\frac{F'}{4\pi}\right|_{r=r_{\text{H}}}=\frac{1}{4\pi}
    \left(
        \frac{2M}{\left(2M-\frac{\alpha}{2M}\right)^2}
        - \frac{4\alpha M^2}{\left(2M-\frac{\alpha}{2M}\right)^5}
    \right).
\end{equation}
For future convenience, we apply a Taylor expansion to $1/T_{\text{H}}$, which yields
\begin{equation}\label{eq:Tay_T-1}
    T_{\text{H}}^{-1}=8\pi M+\frac{\pi \alpha}{M}+O(\alpha^{2}).
\end{equation}
To compute the energy emission rate  of Hawking radiation with the spectrum \eqref{eq:eq:haw_spe}, we need to compute the greybody factor $T_l(\o)$.
In this work, we follow the approach developed in Ref. \cite{harmark2010greybody}.
From the effective potential introduced in Eq. \eqref{eq:def:eff_potential}, we find that its peak increases with the multipole index $l$.
As a result, low frequency with large $l$ are mostly predominantly reflected by the potential $V_l(r)$ and can be safely neglected (see also Ref. \cite{belfaqih2025hawking}). 
Therefore, in the present analysis, we restrict our attention to the $0$-mode, for which the corresponding effective potential $V_0(r)$ is given by 
\begin{equation}
    V_0(r)=-(
    1
    -\frac{2M}{r}
    +\frac{\alpha M^2}{r^4}
)(\frac{4\alpha M^2}{r^6}
-\frac{2M}{r^3}).
\end{equation}
In this approach, the spacetime is divided into the following three regions:
\begin{itemize}
    \item Region I: Near the event horizon $r \approx r_{\text{H}}$, with $V_0(r)\ll \o^2$.
    \item Region III: The asymptotic region, given by $r\gg r_{\text{H}}$.
    \item Region II: the intermediate region between region I and III, defined by $V_0(r)\gg \o^2$.
\end{itemize}
With the spacetime regions introduced above, one can solve the EOM \eqref{eq:EOM_non_tortoise} by requiring the continuity of the solutions across the boundaries.
As a result, the greybody factor corresponding the $l=0$ mode yields
\begin{equation}
    T_0(\o)=\frac{J_{\text{H}}}{J_{in}}=4\o^2r_{\text{H}}^2,
\end{equation}
where $J_{in}$ and $J_{\text{H}}$ are the currents corresponding to the incident and transmitted waves, respectively.
With Eq. \eqref{eq:hor_rad}, we find
\begin{equation}
    T_0(\o)=4\o^2(2M-\frac{\alpha}{8M})^2.
\end{equation}
The energy emission rate of the Hawking radiation is given by \cite{belfaqih2025hawking}
\begin{equation}
    \frac{\dd M}{\dd t}
    =-\frac{1}{2\pi}\int_0^\infty\dd \o\frac{\o T_{0}(\o)}{\exp(\o/T_{\text{H}})-1}.
\end{equation}
Next, with Eq. \eqref{eq:Tay_T-1}, we find 
\begin{equation}
    \frac{\dd M}{\dd t}
    =-\frac{1}{7680\pi M^2}
    +\frac{3\alpha }{20480\pi M^4}+O(\alpha^{2}).
\end{equation}
In Fig \ref{fig:minimal_rate}, we plot the energy emission rate of the Hawking radiation as a function of the BH mass $M$, comparing the classical Schwarzschild BH with the q-OS BH.

    Our findings for the q-OS BH are similar to those for the holonomy-corrected BH analyzed in Ref. \cite{belfaqih2025hawking}.
    In the classical  regime, where $M\gg \alpha$, the energy emission rate of Hawking radiation for the q-OS behaves similarly to that of a classical Schwarzschild BH.
    However, in the Planck regime, characterized by $M^2\sim \alpha$, the energy emission rates of these two types of BHs differ significantly.
    For the classical Schwarzschild BH, classical GR are assumed to hold throughout the entire evaporation process, so the energy emission rate accelerates as the BH mass decreases.
    Therefore, the classical Schwarzschild BH evaporates completely, leaving no remnant. 
    In contrast, for the q-OS BH, the correction term $\frac{3\alpha }{20480\pi M^4}$ to the energy emission rate arising from the LQG correction plays a crucial role at the late stage of the Hawking radiation process.
    Due to the existence of this LQG correction term, the energy emission rate begins to slow down at $M_r=\frac{3\sqrt{\alpha}}{2}$ and eventually stops evaporating at $M_{\text{final}}=\frac{3}{2\sqrt{2}}\sqrt{\alpha}$.

    According to Ref. \cite{lewandowski2023quantum}, there exists a minimal mass for the formation of an event horizon of the q-OS BH, which is given by $M_{\text{min}}=\frac{4}{3\sqrt{3}}\sqrt{\alpha}$.
    Accordingly, we find that $M_{\text{min}}<M_{\text{final}}$.
    Furthermore, there is a minimal radius of the q-OS BH, which is given by the solution of $-2M/r-\alpha M^2/r^4=0$ \cite{lewandowski2023quantum},
    The corresponding result is 
    \begin{equation}
        r_{\text{b}}
        =\left(\frac{\alpha M}{2}\right)^{\frac{1}{3}}.
    \end{equation}
    During the collapse process, the collapsing matter reaches $r_{\text{b}}$, and then bounces.
    In the later stages of BH evaporation, the quantum fluctuations of the event horizon cannot be neglected, and the area of the event horizon can only be determined up to a quantum uncertainty.
    The BH evaporation  stops when the radius of the event horizon reaches the minimal radius $r_{\text{b}}$.
    As demonstrated in Ref. \cite{parikh2025quantum}, the BH horizon area fluctuation is given by $r_{\text{H}}\sqrt{8\pi\alpha}$.
    As following the strategy in \cite{belfaqih2025hawking}, we have replaced the Planck length by $\sqrt{\alpha}$ as the fundamental UV-scalar for the q-OS BH.
    This assumption is justified by two facts: i) $\alpha$ has the dimension of area, which is proportional to the Planck length $\ell_p^2$, see \cite{lewandowski2023quantum}; 
    $\alpha$ encodes the quantum gravity effect of the q-OS BH.
     Then, the BH mass when the BH radius reaches the minimal radius $r_{\text{b}}$ is determined by the following equation
     \begin{equation}
        \mathscr{A}-r_{\text{H}}\sqrt{8\pi\alpha}=\pi r_{\text{b}}^2.
     \end{equation}

\begin{figure}[H] 
	\centering 
	\includegraphics[width=0.4\textwidth]{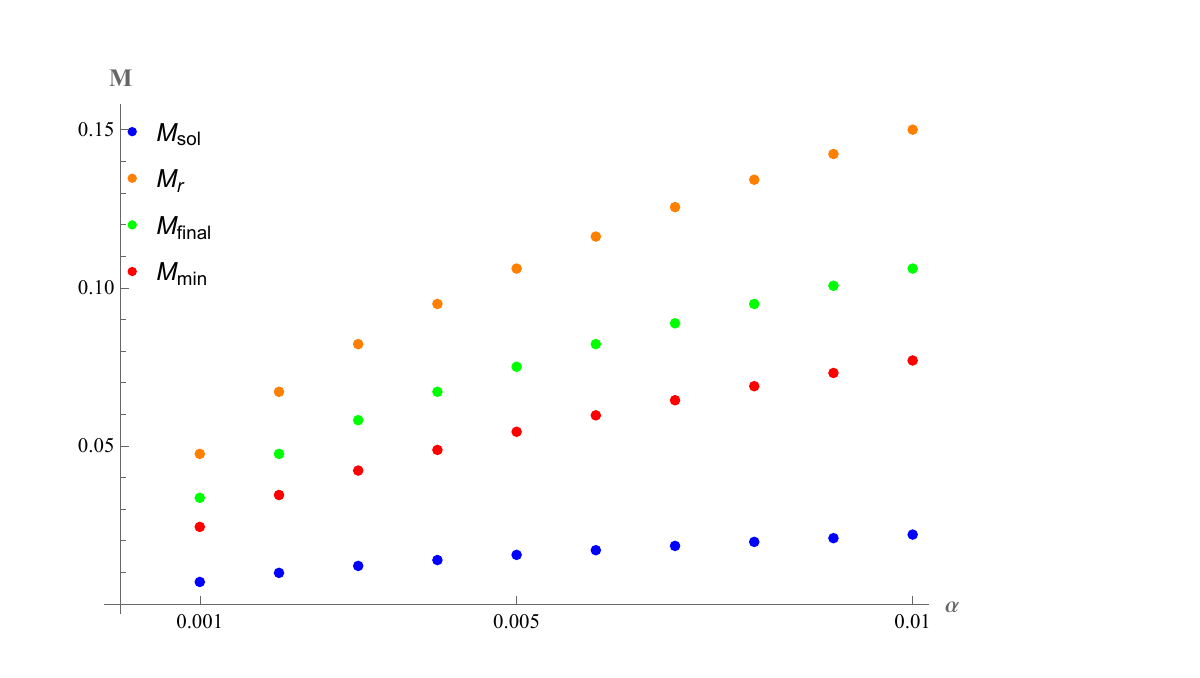} 
	\caption{$\alpha \in[0.001,0.01]$. The blue dots represent $M_{\text{sol}}$. The orange dots represent $M_{r}$. The green dots represent $M_{\text{final}}$. The red dots represent $M_{\text{min}}$.} 
	\label{fig:Plot_of_M_sol} 
	\end{figure}

Here, $\mathscr{A}$ is the non-fluctuated area of the BH horizon. We denote the solution for this equation as $M_{\text{sol}}$.
In Fig. \ref{fig:Plot_of_M_sol}, we compare $M_{\text{sol}}$, $M_{\text{min}}$, $M_r$, and $M_{\text{final}}$.
Here, the parameter $\alpha$ is taken to be in the range of $0.001-0.01$.
From Fig. \ref{fig:Plot_of_M_sol}, it can be observed that $M_\text{final}$ is consistently greater than $M_\text{sol}$.
Therefore, the BH evaporation will terminate prior to the dissolution of the BH horizon in our scenario.

\begin{figure}[H] 
	\centering 
	\includegraphics[width=0.4\textwidth]{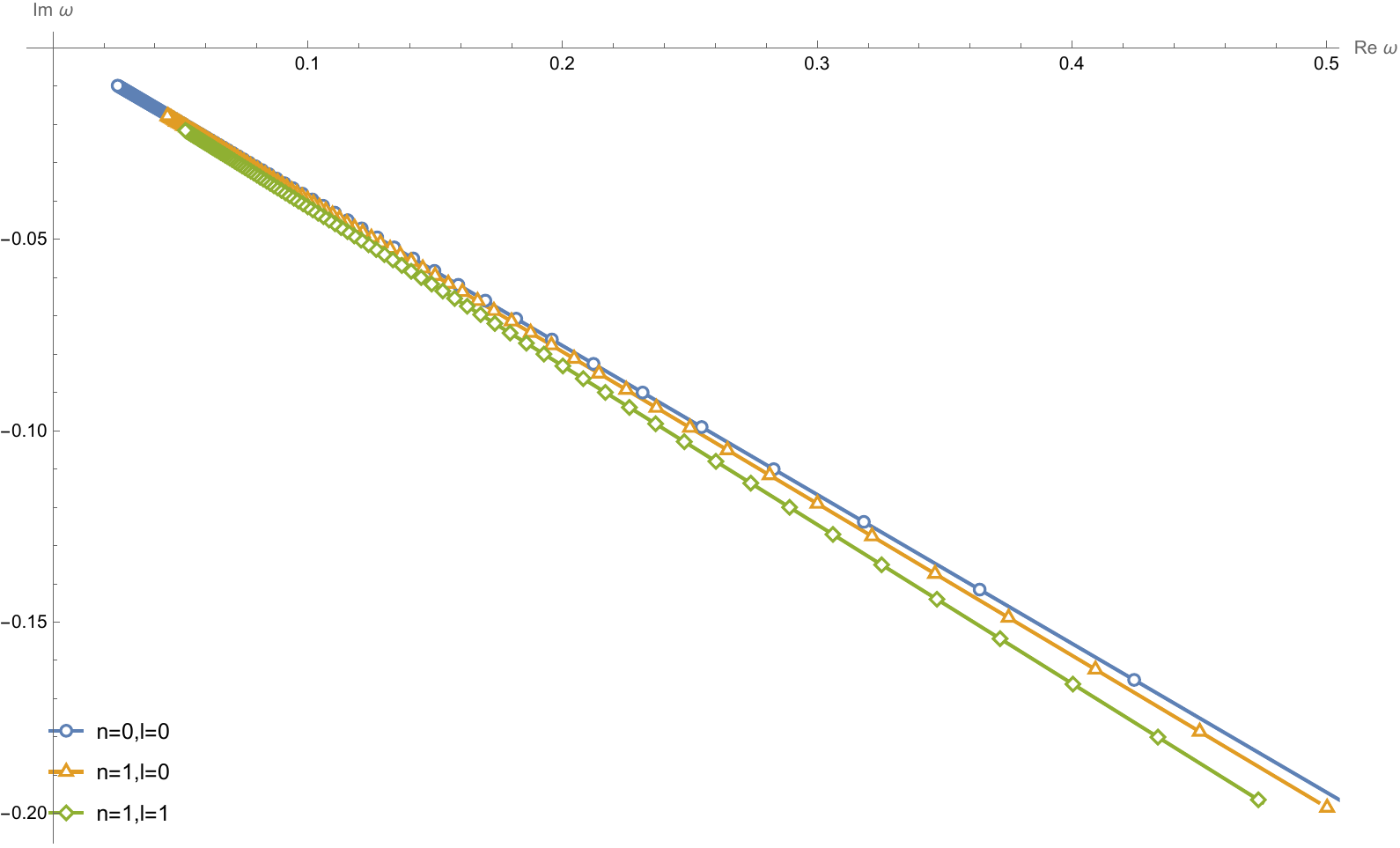} 
	\caption{The QNMs for the q-OS BH with a minimally coupled scalar field. $\alpha=0.001$, $M\in[0.1,10.0]$. } 
	\label{fig:QNM_mini} 
	\end{figure}

To examine the stability of the BH remnant, we need to analyze its QNMs. 
The effective potential $V_l(r)$ in Eq. \eqref{eq:def:eff_potential} enables  the use of the WKB approximation to compute the QNM.
In this paper, we adopt the higher-order WKB approximation introduced in \cite{iyer1987black}.
The numerical results are presented in Fig. \ref{fig:QNM_mini}.
In Fig. \ref{fig:QNM_mini}, we  set $\alpha=0.001$, and $M$ is taken to be the range of $0.1-10.0$.
We have computed the QNM with the modes of $n=0,\,l=0$, $n=1,\,l=0$, and $n=1,\,l=1$.
We find that $\text{Im}\, \o< 0$, which indicates the stability of the BH remnant.

Our results indicate that the information of the collapsing matter is stored in the remnant of the BH, rather than being lost through Hawking radiation.
Our findings suggest that the LQG corrected BH potentially provide a viable framework for understanding the BH information paradox, since the information of the collapsing matter can be stored in the BBH remnant.
\section{Black hole evaporation with non-minimally coupled scalar field}\label{sec:non-min}
\subsection{The equation of motion of the non-minimally coupled scalar field}
For a non-minimally coupled scalar field, its EOM reads 
\begin{equation}
    \nabla_a\nabla^a\phi-\xi g(r) R \phi=0.
\end{equation}
Here, R is the Ricci scalar of the q-OS spacetime, given by $R=-\frac{6\alpha M^2}{r^6}$,
and $\xi$ denotes the coupling constant.
$g(r)$ is a function of the radial coordinate $r$, which satisfies the following requirements: i) $\lim_{r\to r_{\text{H}}}g(r)=\lim_{r\to r_{\text{H}}}F(r)^{-1}$; ii) $g(r)$ is finite when $r$ is significantly larger than $r_{\text{H}}$; iii) $\lim_{r\to\infty}g(r)$=1.
Following a similar approach to that introduced in Sec. \ref{sec:min}, the EOM for $u_{lm}$ in this case becomes 
\begin{equation}\label{eq:nonmin_ulm}
    \frac{\partial^2}{\partial r_*^2}u_{lm}+\left(\o^2+\xi F(r)g(r)\frac{6\alpha M^2}{r^6}-V_l(r)\right)u_{lm}=0.
\end{equation}
Compared to the minimally coupled case, an additional term $\xi \frac{6\alpha M^2}{r^6}$ arises due to the non-minimal coupling.
\subsection{The greybody factor and black hole evaporation}
To compute the greybody in this case, we divide the spacetime into three regions, following the same procedure as in Sec. \ref{sec:min}.

\begin{figure}[H] 
	\centering 
	\includegraphics[width=0.4\textwidth]{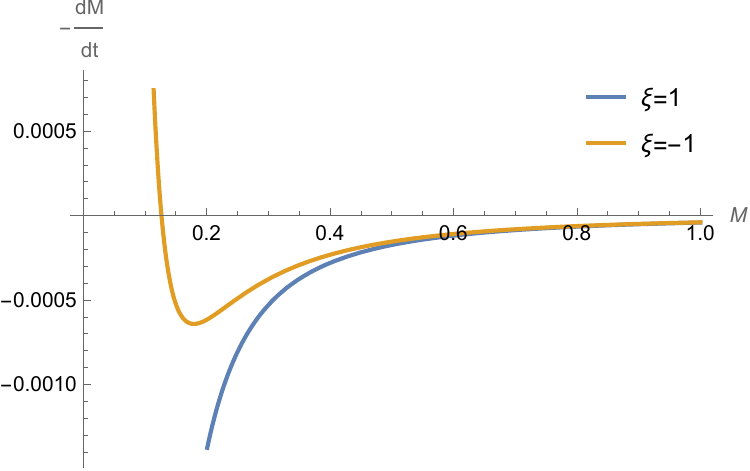} 
	\caption{The energy emission rate of the q-OS BH with non-minimally coupled scalar field. The parameter $\alpha=0.001$. } 
	\label{fig:non_minial_emission_rate} 
	\end{figure}

In the region I, where the potential $V_0(r)$ can be neglected.
Furthermore, the radial coordinate $r$ is approximately near the horizon
\begin{equation}
    r\approx r_{\text{H}}\approx2M-\frac{\alpha}{8M}.
\end{equation}
Therefore, in region I, the EOM \eqref{eq:nonmin_ulm} simplifies to
\begin{equation}
    \frac{\partial^2}{\partial r_*^2}u_0+\left(\o^2+\xi\frac{6\alpha M^2}{\left(2M-\frac{\alpha}{8M}\right)^6}\right)u_0=0.
\end{equation}
Here, as before, we only consider the $l=0$-mode of $u_{lm}$.
Following the techniques of Ref. \cite{harmark2010greybody}, we consider only the purely ingoing wave. We then obtain
\begin{equation}
u_0=A_{\text{I}}\exp\left( ir_*\tilde{\o}\right),
\end{equation}
where $\tilde{\o}=\sqrt{\o^2+\xi\frac{6\alpha M^2}{\left(2M-\frac{\alpha}{8M}\right)^6}}$
Since the term $\xi\frac{6\alpha M^2}{\left(2M-\frac{\alpha}{8M}\right)^6}$ is proportional to the quantum correction parameter $\alpha$, it is natural to assume that 
\begin{equation}
    \o^2\gg\frac{6\alpha M^2}{\left(2M-\frac{\alpha}{8M}\right)^6}.
\end{equation}
We do not neglect this quantum correction in our analysis, because different values of the parameter $\xi$ can lead to qualitatively different outcomes for the fate of the q-OS BH, which we will see later. 
At the region close to the horizon, the function $F(r)$ appears in the line element \eqref{eq:quan_OS} can be expressed as 
\begin{equation}
    F(r)\approx 2\kappa_{\text{H}}
    \left(r-r_{\text{H}}\right),
\end{equation}
with the surface $\kappa_{\text{H}}$ is given by 
\begin{equation}
    \kappa_{\text{H}}:=\frac{1}{2}F'(r)|_{r=r_{\text{H}}}.
\end{equation}
Therefore, using Eq. \eqref{eq:def:tortoise}, the tortoise coordinate in this region approximately yields 
\begin{equation}
    r_*\approx\frac{1}{2\kappa_{\text{H}}}\log\left(\frac{r-r_{\text{H}}}{r_{\text{H}}}\right).
\end{equation}
Then, as argued in Ref. \cite{harmark2010greybody}, we have $\tilde{\o}\approx\o\ll\kappa_{\text{H}}$. Eventually, we find
\begin{equation}\label{eq:sol_for_I}
    u_{0}=A_{\text{I}}
    \left[
        1+i\frac{\tilde{\o}}{2\kappa_{\text{H}}}
        \log\left(
            \frac{r-r_{\text{H}}}{r_{\text{H}}}
        \right)
    \right].
\end{equation}
For the region II, where $V_0(r)\gg \o^2$.
Since Both $F(r)$ and $g(r)$ is finite in this region, the quantum correction $\xi F(r)^{-1}g(r)\frac{6\alpha M^2}{r^6}$ can be neglected.
Therefore, the EOM in this region is reduced to
\begin{equation}
   \partial_r\left(r^2F(r)\partial_ru_0\right)=0.
\end{equation}
The most general solution to this equation is
\begin{equation}
    u_0(r)=A_{\text{II}}+B_{\text{II}}G(r),
\end{equation}
with 
\begin{equation}
    G(r)=\int_{\infty}^r\frac{\dd r'}{r^2F(r)}.
\end{equation}
For the case of $r\approx r_{\text{H}}$, we have 
\begin{equation}
    G(r)\approx\frac{1}{2r^2_{\text{H}}\kappa_{\text{H}}}
    \log\left(\frac{r-r_{\text{H}}}{r_{\text{H}}}\right).
\end{equation}
Since this region is close to region I, the solution can be matched to that in region I.
By comparing with Eq. \eqref{eq:sol_for_I}, we find 
\begin{equation}
    A_{\text{II}}=A_{\text{I}},\,
    B_{\text{II}}=i\tilde{\o}r^2_{\text{H}}A_{\text{I}}.
\end{equation}
At the part close to region III, where $r\gg r_{\text{H}}$, we have $F(r)\approx 1$. Therefore,
\begin{equation}
    G(r)\approx\int_{\infty}^{r}\frac{\dd r}{r^2}=-\frac{1}{2r}.
\end{equation}
The solution of EOM at this part takes the following form
\begin{equation}\label{eq:sol_reg_II-III}
    u_{0}
    =A_{\text{I}}
    \left(
        1-i\tilde{\o}r^2_{\text{H}}
        \frac{1}{2r}
    \right).
\end{equation}
Region III is the asymptotically flat region, where the solution of the EOM in is given in \cite{harmark2010greybody}
\begin{equation}\label{eq:solu_reg_III}
u_0=\sqrt{\frac{2}{\pi}}\rho^{-1}
\left(
    iC_1\exp(-i\rho)
    -iC_2\exp(i\rho)
\right),
\end{equation}
with $\rho=\tilde{\o}r$.
Here, we have replaced $\o$ by $\tilde{\o}$ to match the solution in region II.
Near region II, the low-frequency limit implies $\rho\ll1$.
Consequently, Eq. \eqref{eq:solu_reg_III} can be expanded as 
\begin{equation}
    u_0=\sqrt{\frac{2}{\pi}}
    \left(
        C_1+C_i
        -i\frac{C_2-C_1}{\rho}
    \right)
    +O(\rho).
\end{equation}
Then, comparing this result with Eq. \eqref{eq:sol_reg_II-III}, we obtain
\begin{equation}
    C_1+C_2=\sqrt{\frac{\pi}{2}}A_{\text{I}},\quad
    C_2-C_1=\sqrt{\frac{\pi}{2}}\tilde{\o}^2r^2_{\text{H}}A_{\text{I}}.
\end{equation}
In the low-frequency limit, we find 
\begin{equation}
    \frac{|C_2-C_1|}{|C_1+C_2|}\ll 1.
\end{equation}
Finally, following the procedure of \cite{harmark2010greybody}, the greybody factor is given by
\begin{equation}
    T_0(\o)
    =4\pi\tilde{\o}^2r{_\text{H}}^2.
\end{equation}
Therefore, the energy emission rate of the Hawking radiation is computed as 
\begin{equation}
    \frac{\dd M}{\dd t}
    =-\frac{1}{2\pi}\int_0^\infty\dd \o\frac{\o T_{0}(\o)}{\exp(\o/T_H)-1}
    =-\frac{8M^2-9\alpha+120\xi\alpha}{61400\pi M^4}+O(\alpha^{2}).
\end{equation}
Here, we find that in the non-minimal coupling case, the values of the coupling constant $\xi$ affect behavior of Hawking radiation significantly at the late stages,  potentially leading to different fate of the BHs.
Fig \ref{fig:non_minial_emission_rate} presents the energy emission rate of the Hawking radiation with the cases of $\xi=\pm1$.
    From Fig. \ref{fig:non_minial_emission_rate}, we learn that for $\xi=1$, the behavior of Hawking radiation is very similar to that of the classical Schwarzschild BH. In this case, the quantum gravity correction does not slow down the emission process, and the black hole evaporates completely if no additional mechanisms are taken into account.
    In contrast, for $\xi=-1$, the behavior of Hawking radiation closely resembles that of the minimally coupled scalar field in the q-OS BH.
    The evaporation begins to slow down at $M'_r=\frac{\sqrt{129\alpha}}{2}$ and terminates at $M'_{\text{final}}=\frac{\sqrt{129\alpha}}{2\sqrt{2}}$.
    This case therefore also indicates the existence of a BH remnant.
    Furthermore, since $M'_{\text{final}}>M_{\text{final}}$, 
    the BH evaporation in the $\xi=-1$ case stops before the dissolution of the BH horizon.

\textit{Remark:} When $\xi>0$, we find that for $\xi=0.075$, the energy emission rate coincides with that of the classical Schwarzschild BH.
When $\xi$ is very close to slightly smaller than 0.075, the non-minimal coupling effects may lead to the formation of a BH remnant;
one representative  example is $\xi=0.074$. However, in this case, $M'_r\approx 0.17\sqrt{\alpha}$.
At such a scale, the semi-classical analysis adopted above may no longer be valid, since $\alpha\gg M^2$.
Quantum fluctuations are expected to become significant, and additional subtleties may arise beyond the present framework.
Therefore, in the remainder of this paper, we focus on the case $\xi=-1$, while leaving a detailed investigation of these subtleties for future work.

\begin{figure}[H] 
	\centering 
	\includegraphics[width=0.4\textwidth]{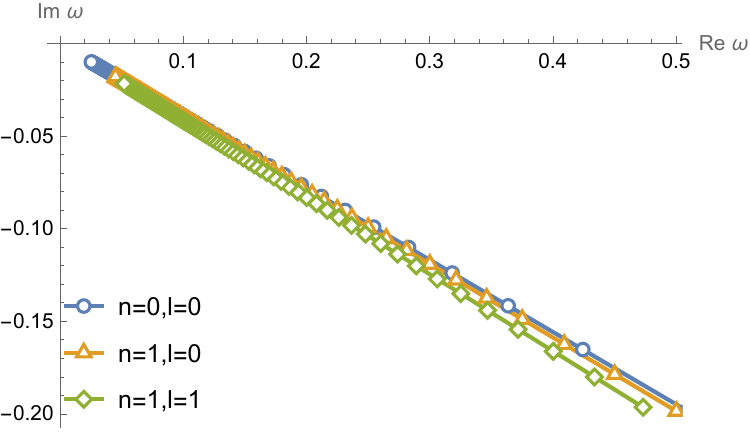} 
	\caption{The QNMs of the q-OS BH with  non-minimal scalar with the case of $\xi=-1$. $\alpha=0.001$, $M\in[0.1,10.0]$. } 
	\label{fig:plot_nonminimal_QNM} 
	\end{figure}

We then analyze the QNM of the BH for the case of $\xi=-1$ using higher-order WKB approach.
To compute the QNMs, the specific form of $g(r)$need to be specified.
Without lose of generality, we choose $g(r)=F(r)^{-1}$.
Then the results of the QNMs are shown in Fig \ref{fig:plot_nonminimal_QNM}, which also indicate that $\text{Im}\, \o< 0$, thereby implying the stability of the BH remnant.
    \textit{Remark:} Indeed, the specific form of $g(r)$ does not affect the stability of the BH remnant, as long as it satisfies the requirements introduced above.
    a correction term $-F(r)^{-1}g(r)\frac{6\alpha M^2}{r^6}$, arising from the non-minimal coupling effect, is proportional to the quantum parameter $\alpha$.
    Hence, isn the region where $r\gg r_{H}$, the different choices of $g(r)$ only induce very slight changes to the shape of $V_0(r)$.
    The WKB approach is not sensitive to such small modifications of the potential, leading to similar results for different choices of $g(r)$.
    \section{Conclusions and outlooks}\label{sec:con}
In this paper, we investigate the evaporation process of the q-OS BH within a semiclassical framework.
We consider both the minimally coupled and non-minimally coupled scalar fields as the test fields.
For the minimally coupled case, we find that quantum gravity corrections play a crucial role at the late stage of the evaporation process. The evaporation rate first slows down and then eventually terminates. Further analysis shows that the evaporation process stops before the dissolution of the event horizon. Moreover, the QNM analysis indicates the stability of the resulting BH remnant.
For the non-minimally coupled case, different values of the coupling parameter $\xi$ play a crucial role in determining the fate of the BH. In this work, we focus on the cases $\xi = \pm 1$. For $\xi = 1$, the energy emission rate of Hawking radiation behaves very similarly to that of the classical Schwarzschild BH, and the evaporation process accelerates significantly at the late stage. In contrast, for $\xi = -1$, the evaporation process also slows down and eventually terminates at late times.

Our findings imply that the LQG effects potentially provide a resolution of the BH information paradox.
The quantum correction term $\frac{\alpha M^2}{r^4}$ becomes significant at the late stage of the BH evaporation process, leading to the formation of a BH remnant.
As a result, the information of the collapsing matter is expected to be stored in this remnant, which may (at least partially) resolve the BH information paradox. Nevertheless, several difficulties and subtleties remain unresolved, and we leave these issues for future investigations:
\begin{itemize}
    \item In this work, we follow the assumption in Ref. \cite{shi2024higher} that the BH temperature is $T_{\text{H}}=\left.\frac{F'}{4\pi}\right|_{r=r_{\text{H}}}$, known as the Hawking temperature.
    Nevertheless, Hawking radiation is originally  derived based on the quantum field theory in curved spacetime, where the spacetime background is treated as classical.
    In the case of the q-OS BH, However, quantum gravity effects are already taken into account, and the spacetime is therefore no longer purely classical.
    As a result, additional quantum corrections to the BH temperature formula may arise. However, such quantum corrections are expected to be small and should not significantly affect the main results presented in this paper. We leave a detailed investigation of this intriguing issue for future work.
    \item  For the non-minimally coupled cases,  when the value of $\xi$ is very close but slightly smaller than $0.075$, the BH evaporation process appears, at the first glance, slow down and eventually terminate.
    However, in this regime one has $M^2\ll\alpha$, as demonstrated previously. In this Planck-scale region, quantum fluctuations cannot be neglected, and therefore the semiclassical analysis employed in this paper may no longer be valid. To achieve a deeper understanding of the physics in this regime, a more comprehensive treatment of quantum gravity effects is required, which we leave for future investigations.
    To enhance our understanding of the physics in this Planck region, a more detailed discussion on the quantum gravity effects is need in the future researches.
    \item  Our results suggest the formation of a remnant in the evaporation process of the q-OS BH, which may provide new insights into the BH information paradox, as the information of the collapsing matter could be stored in the BH remnant. Nevertheless, the underlying mechanism for resolving the BH information paradox remains unclear. More detailed investigations are required to clarify how information is preserved and recovered during the evaporation process. In particular, demonstrating that the evaporation process is unitary is essential for resolving the information paradox. In this context, the BH purification strategy may offer a useful framework for addressing this issue \cite{umemoto2018entanglement,engelhardt2022canonical}.
\end{itemize}
\section*{Acknowledgments}
HT acknowledges the valuable discussions with Feiyi Liu and Yang Wang. The authors acknowledge the support from the Hunan Provincial Natural Science Foundation of China (Grant No.2022JJ30220).

\bibliographystyle{jhep}
\bibliography{myref}

@article{LIGOScientific:2016aoc,
    author = "Abbott, B. P. and others",
    collaboration = "LIGO Scientific, Virgo",
    title = "{Observation of Gravitational Waves from a Binary Black Hole Merger}",
    eprint = "1602.03837",
    archivePrefix = "arXiv",
    primaryClass = "gr-qc",
    reportNumber = "LIGO-P150914",
    doi = "10.1103/PhysRevLett.116.061102",
    journal = "Phys. Rev. Lett.",
    volume = "116",
    number = "6",
    pages = "061102",
    year = "2016"
}

@article{LIGOScientific:2017bnn,
    author = "Abbott, Benjamin P. and others",
    collaboration = "LIGO Scientific, VIRGO",
    title = "{GW170104: Observation of a 50-Solar-Mass Binary Black Hole Coalescence at Redshift 0.2}",
    eprint = "1706.01812",
    archivePrefix = "arXiv",
    primaryClass = "gr-qc",
    reportNumber = "LIGO-P170104",
    doi = "10.1103/PhysRevLett.118.221101",
    journal = "Phys. Rev. Lett.",
    volume = "118",
    number = "22",
    pages = "221101",
    year = "2017",
    note = "[Erratum: Phys.Rev.Lett. 121, 129901 (2018)]"
}

@article{LIGOScientific:2016dsl,
    author = "Abbott, B. P. and others",
    collaboration = "LIGO Scientific, Virgo",
    title = "{Binary Black Hole Mergers in the first Advanced LIGO Observing Run}",
    eprint = "1606.04856",
    archivePrefix = "arXiv",
    primaryClass = "gr-qc",
    reportNumber = "LIGO-P1600088",
    doi = "10.1103/PhysRevX.6.041015",
    journal = "Phys. Rev. X",
    volume = "6",
    number = "4",
    pages = "041015",
    year = "2016",
    note = "[Erratum: Phys.Rev.X 8, 039903 (2018)]"
}

@article{LIGOScientific:2016sjg,
    author = "Abbott, B. P. and others",
    collaboration = "LIGO Scientific, Virgo",
    title = "{GW151226: Observation of Gravitational Waves from a 22-Solar-Mass Binary Black Hole Coalescence}",
    eprint = "1606.04855",
    archivePrefix = "arXiv",
    primaryClass = "gr-qc",
    reportNumber = "LIGO-P151226",
    doi = "10.1103/PhysRevLett.116.241103",
    journal = "Phys. Rev. Lett.",
    volume = "116",
    number = "24",
    pages = "241103",
    year = "2016"
}

@book{wald2010general,
  title={General relativity},
  author={Wald, Robert M},
  year={2010},
  publisher={University of Chicago press}
}

@book{rovelli2004quantum,
  title={Quantum gravity},
  author={Rovelli, Carlo},
  year={2004},
  publisher={Cambridge university press}
}

@book{thiemann2008modern,
  title={Modern canonical quantum general relativity},
  author={Thiemann, Thomas},
  year={2008},
  publisher={Cambridge University Press}
}

@article{lewandowski2023quantum,
  title={Quantum oppenheimer-snyder and swiss cheese models},
  author={Lewandowski, Jerzy and Ma, Yongge and Yang, Jinsong and Zhang, Cong},
  journal={Physical Review Letters},
  volume={130},
  number={10},
  pages={101501},
  year={2023},
  publisher={APS}
}

@article{yang2023shadow,
  title={Shadow and stability of quantum-corrected black holes},
  author={Yang, Jinsong and Zhang, Cong and Ma, Yongge},
  journal={The European Physical Journal C},
  volume={83},
  number={7},
  pages={619},
  year={2023},
  publisher={Springer}
}

@article{oppenheimer1939continued,
  title={On continued gravitational contraction},
  author={Oppenheimer, J Robert and Snyder, Hartland},
  journal={Physical Review},
  volume={56},
  number={5},
  pages={455},
  year={1939},
  publisher={APS}
}

@book{poisson2004relativist,
  title={A relativist's toolkit: the mathematics of black-hole mechanics},
  author={Poisson, Eric},
  year={2004},
  publisher={Cambridge university press}
}

@article{ashtekar2006quantum,
  title={Quantum nature of the big bang},
  author={Ashtekar, Abhay and Pawlowski, Tomasz and Singh, Parampreet},
  journal={Physical review letters},
  volume={96},
  number={14},
  pages={141301},
  year={2006},
  publisher={APS}
}

@article{parikh2000hawking,
  title={Hawking radiation as tunneling},
  author={Parikh, Maulik K and Wilczek, Frank},
  journal={Physical review letters},
  volume={85},
  number={24},
  pages={5042},
  year={2000},
  publisher={APS}
}

@article{parikh2004secret,
  title={A secret tunnel through the horizon},
  author={Parikh, Maulik},
  journal={International Journal of Modern Physics D},
  volume={13},
  number={10},
  pages={2351--2354},
  year={2004},
  publisher={World Scientific}
}

@article{zhang2009black,
  title={Black hole entropy, log correction and inverse area correction},
  author={Zhang, Jingyi},
  journal={Physics Letters B},
  volume={675},
  number={1},
  pages={14--17},
  year={2009},
  publisher={Elsevier}
}

@article{zhang2008black,
  title={Black hole quantum tunnelling and black hole entropy correction},
  author={Zhang, Jingyi},
  journal={Physics Letters B},
  volume={668},
  number={5},
  pages={353--356},
  year={2008},
  publisher={Elsevier}
}

@article{event2019first,
  title={First M87 event horizon telescope results. I. The shadow of the supermassive black hole},
  author={Event Horizon Telescope Collaboration and others},
  journal={arXiv preprint arXiv:1906.11238},
  year={2019}
}

@article{akiyama2019first,
  title={First M87 event horizon telescope results. II. Array and instrumentation},
  author={Akiyama, Kazunori and Alberdi, Antxon and Alef, Walter and Asada, Keiichi and Azulay, Rebecca and Baczko, Anne-Kathrin and Ball, David and Balokovi{\'c}, Mislav and Barrett, John and Bintley, Dan and others},
  journal={The Astrophysical Journal Letters},
  volume={875},
  number={1},
  pages={L2},
  year={2019},
  publisher={IOP Publishing}
}

@article{akiyama2019first-2,
  title={First M87 event horizon telescope results. III. Data processing and calibration},
  author={Akiyama, Kazunori and Alberdi, Antxon and Alef, Walter and Asada, Keiichi and Azulay, Rebecca and Baczko, Anne-Kathrin and Ball, David and Balokovi{\'c}, Mislav and Barrett, John and Bintley, Dan and others},
  journal={The Astrophysical Journal Letters},
  volume={875},
  number={1},
  pages={L3},
  year={2019},
  publisher={IOP Publishing}
}

@article{akiyama2019first-3,
  title={First M87 event horizon telescope results. IV. Imaging the central supermassive black hole},
  author={Akiyama, Kazunori and Alberdi, Antxon and Alef, Walter and Asada, Keiichi and Azulay, Rebecca and Baczko, Anne-Kathrin and Ball, David and Balokovi{\'c}, Mislav and Barrett, John and Bintley, Dan and others},
  journal={The Astrophysical Journal Letters},
  volume={875},
  number={1},
  pages={L4},
  year={2019},
  publisher={IOP Publishing}
}

@article{akiyama2019first-4,
  title={First M87 event horizon telescope results. V. Physical origin of the asymmetric ring},
  author={Akiyama, Kazunori and Alberdi, Antxon and Alef, Walter and Asada, Keiichi and Azulay, Rebecca and Baczko, Anne-Kathrin and Ball, David and Balokovi{\'c}, Mislav and Barrett, John and Bintley, Dan and others},
  journal={The Astrophysical Journal Letters},
  volume={875},
  number={1},
  pages={L5},
  year={2019},
  publisher={IOP Publishing}
}

@article{akiyama2019first-5,
  title={First M87 event horizon telescope results. VI. The shadow and mass of the central black hole},
  author={Akiyama, Kazunori and Alberdi, Antxon and Alef, Walter and Asada, Keiichi and Azulay, Rebecca and Baczko, Anne-Kathrin and Ball, David and Balokovi{\'c}, Mislav and Barrett, John and Bintley, Dan and others},
  journal={The Astrophysical Journal Letters},
  volume={875},
  number={1},
  pages={L6},
  year={2019},
  publisher={IOP Publishing}
}

@article{abbott2016observation,
  title={Observation of gravitational waves from a binary black hole merger},
  author={Abbott, Benjamin P and Abbott, Richard and Abbott, TDe and Abernathy, MR and Acernese, Fausto and Ackley, Kendall and Adams, Carl and Adams, Thomas and Addesso, Paolo and Adhikari, Rana X and others},
  journal={Physical review letters},
  volume={116},
  number={6},
  pages={061102},
  year={2016},
  publisher={APS}
}

@article{han2007fundamental,
  title={Fundamental structure of loop quantum gravity},
  author={Han, Muxin and Ma, Yongge and Huang, Weiming},
  journal={International Journal of Modern Physics D},
  volume={16},
  number={09},
  pages={1397--1474},
  year={2007},
  publisher={World Scientific}
}

@incollection{thiemann2003lectures,
  title={Lectures on loop quantum gravity},
  author={Thiemann, Thomas},
  booktitle={Quantum gravity: From theory to experimental search},
  pages={41--135},
  year={2003},
  publisher={Springer}
}

@article{ashtekar2004background,
  title={Background independent quantum gravity: a status report},
  author={Ashtekar, Abhay and Lewandowski, Jerzy},
  journal={Classical and Quantum Gravity},
  volume={21},
  number={15},
  pages={R53},
  year={2004},
  publisher={IOP Publishing}
}

@article{giesel2012classical,
  title={From classical to quantum gravity: introduction to loop quantum gravity},
  author={Giesel, Kristina and Sahlmann, Hanno},
  journal={arXiv preprint arXiv:1203.2733},
  year={2012}
}

@article{rovelli2011zakopane,
  title={Zakopane lectures on loop gravity},
  author={Rovelli, Carlo},
  journal={arXiv preprint arXiv:1102.3660},
  year={2011}
}

@article{perez2013spin,
  title={The spin-foam approach to quantum gravity},
  author={Perez, Alejandro},
  journal={Living Reviews in Relativity},
  volume={16},
  pages={1--128},
  year={2013},
  publisher={Springer}
}

@article{haggard2015quantum,
  title={Quantum-gravity effects outside the horizon spark black to white hole tunneling},
  author={Haggard, Hal M and Rovelli, Carlo},
  journal={Physical Review D},
  volume={92},
  number={10},
  pages={104020},
  year={2015},
  publisher={APS}
}

@article{bianchi2018white,
  title={White holes as remnants: a surprising scenario for the end of a black hole},
  author={Bianchi, Eugenio and Christodoulou, Marios and d’Ambrosio, Fabio and Haggard, Hal M and Rovelli, Carlo},
  journal={Classical and Quantum Gravity},
  volume={35},
  number={22},
  pages={225003},
  year={2018},
  publisher={IOP Publishing}
}

@article{hawking1975particle,
  title={Particle creation by black holes},
  author={Hawking, Stephen W},
  journal={Communications in mathematical physics},
  volume={43},
  number={3},
  pages={199--220},
  year={1975},
  publisher={Springer}
}

@article{almheiri2021entropy,
  title={The entropy of Hawking radiation},
  author={Almheiri, Ahmed and Hartman, Thomas and Maldacena, Juan and Shaghoulian, Edgar and Tajdini, Amirhossein},
  journal={Reviews of Modern Physics},
  volume={93},
  number={3},
  pages={035002},
  year={2021},
  publisher={APS}
}

@article{ghosh2005log,
  title={Log correction to the black hole area law},
  author={Ghosh, Amit and Mitra, Parthasarathi},
  journal={Physical Review D—Particles, Fields, Gravitation, and Cosmology},
  volume={71},
  number={2},
  pages={027502},
  year={2005},
  publisher={APS}
}

@article{page1993information,
  title={Information in black hole radiation},
  author={Page, Don N},
  journal={Physical review letters},
  volume={71},
  number={23},
  pages={3743},
  year={1993},
  publisher={APS}
}

@article{page2013time,
  title={Time dependence of Hawking radiation entropy},
  author={Page, Don N},
  journal={Journal of Cosmology and Astroparticle Physics},
  volume={2013},
  number={09},
  pages={028},
  year={2013},
  publisher={IOP Publishing}
}

@article{kaul2000logarithmic,
  title={Logarithmic correction to the Bekenstein-Hawking entropy},
  author={Kaul, Romesh K and Majumdar, Parthasarathi},
  journal={Physical Review Letters},
  volume={84},
  number={23},
  pages={5255},
  year={2000},
  publisher={APS}
}

@article{domagala2004black,
  title={Black-hole entropy from quantum geometry},
  author={Domagala, Marcin and Lewandowski, Jerzy},
  journal={Classical and Quantum Gravity},
  volume={21},
  number={22},
  pages={5233},
  year={2004},
  publisher={IOP Publishing}
}

@article{meissner2004black,
  title={Black-hole entropy in loop quantum gravity},
  author={Meissner, Krzysztof A},
  journal={Classical and Quantum Gravity},
  volume={21},
  number={22},
  pages={5245},
  year={2004},
  publisher={IOP Publishing}
}

@book{frolov2011introduction,
  title={Introduction to black hole physics},
  author={Frolov, Valeri P and Zelnikov, Andrei},
  year={2011},
  publisher={OUP Oxford}
}

@article{afshordi2024black,
  title={Black Holes Inside and Out 2024: visions for the future of black hole physics},
  author={Afshordi, Niayesh and Ashtekar, Abhay and Barausse, Enrico and Berti, Emanuele and Brito, Richard and Buoninfante, Luca and Carballo-Rubio, Ra{\'u}l and Cardoso, Vitor and Carullo, Gregorio and Dafermos, Mihalis and others},
  journal={arXiv preprint arXiv:2410.14414},
  year={2024}
}

@misc{chandrasekhar1985mathematical,
  title={The mathematical theory of black holes},
  author={Chandrasekhar, Subrahmanyan and Thorne, Kip S},
  year={1985},
  publisher={American Association of Physics Teachers}
}

@article{stashko2024quasinormal,
  title={Quasinormal modes and gray-body factors of regular black holes in asymptotically safe gravity},
  author={Stashko, Oleksandr},
  journal={Physical Review D},
  volume={110},
  number={8},
  pages={084016},
  year={2024},
  publisher={APS}
}

@article{zhang2023black,
  title={Black hole image encoding quantum gravity information},
  author={Zhang, Cong and Ma, Yongge and Yang, Jinsong},
  journal={Physical Review D},
  volume={108},
  number={10},
  pages={104004},
  year={2023},
  publisher={APS}
}

@article{gong2024quasinormal,
  title={Quasinormal modes of quantum-corrected black holes},
  author={Gong, Huajie and Li, Shulan and Zhang, Dan and Fu, Guoyang and Wu, Jian-Pin},
  journal={Physical Review D},
  volume={110},
  number={4},
  pages={044040},
  year={2024},
  publisher={APS}
}

@article{yang2024gravitational,
  title={Gravitational waveforms from periodic orbits around a quantum-corrected black hole},
  author={Yang, Sen and Zhang, Yu-Peng and Zhu, Tao and Zhao, Li and Liu, Yu-Xiao},
  journal={arXiv preprint arXiv:2407.00283},
  year={2024}
}

@article{liu2024gravitational,
  title={Gravitational lensing effect of black holes in effective quantum gravity},
  author={Liu, Hao and Lai, Meng-Yun and Pan, Xiao-Yin and Huang, Hyat and Zou, De-Cheng},
  journal={Physical Review D},
  volume={110},
  number={10},
  pages={104039},
  year={2024},
  publisher={APS}
}

@article{chatterjee2004universal,
  title={Universal canonical black hole entropy},
  author={Chatterjee, Ashok and Majumdar, Parthasarathi},
  journal={Physical Review Letters},
  volume={92},
  number={14},
  pages={141301},
  year={2004},
  publisher={APS}
}

@article{medved2004comment,
  title={A comment on black hole entropy or does nature abhor a logarithm?},
  author={Medved, AJM},
  journal={Classical and Quantum Gravity},
  volume={22},
  number={1},
  pages={133},
  year={2004},
  publisher={IOP Publishing}
}

@article{lin2024effective,
  title={Effective four-dimensional loop quantum black hole with a cosmological constant},
  author={Lin, Jianhui and Zhang, Xiangdong},
  journal={Physical Review D},
  volume={110},
  number={2},
  pages={026002},
  year={2024},
  publisher={APS}
}

@article{shi2024higher,
  title={Higher-dimensional quantum Oppenheimer-Snyder model},
  author={Shi, Zijian and Zhang, Xiangdong and Ma, Yongge},
  journal={Physical Review D},
  volume={110},
  number={10},
  pages={104074},
  year={2024},
  publisher={APS}
}

@article{banerjee2008quantum,
  title={Quantum tunneling and back reaction},
  author={Banerjee, Rabin and Majhi, Bibhas Ranjan},
  journal={Physics Letters B},
  volume={662},
  number={1},
  pages={62--65},
  year={2008},
  publisher={Elsevier}
}

@article{banerjee2008quantum2,
  title={Quantum tunneling beyond semiclassical approximation},
  author={Banerjee, Rabin and Majhi, Bibhas Ranjan},
  journal={Journal of High Energy Physics},
  volume={2008},
  number={06},
  pages={095},
  year={2008},
  publisher={IOP Publishing}
}

@article{banerjee2009quantum,
  title={Quantum tunneling and trace anomaly},
  author={Banerjee, Rabin and Majhi, Bibhas Ranjan},
  journal={Physics Letters B},
  volume={674},
  number={3},
  pages={218--222},
  year={2009},
  publisher={Elsevier}
}

@article{majhi2009fermion,
  title={Fermion tunneling beyond semiclassical approximation},
  author={Majhi, Bibhas Ranjan},
  journal={Physical Review D—Particles, Fields, Gravitation, and Cosmology},
  volume={79},
  number={4},
  pages={044005},
  year={2009},
  publisher={APS}
}

@article{majhi2010hawking,
  title={Hawking radiation due to photon and gravitino tunneling},
  author={Majhi, Bibhas Ranjan and Samanta, Saurav},
  journal={Annals of Physics},
  volume={325},
  number={11},
  pages={2410--2424},
  year={2010},
  publisher={Elsevier}
}

@article{zi2024eccentric,
  title={Eccentric extreme mass-ratio inspirals: A gateway to probe quantum gravity effects},
  author={Zi, Tieguang and Kumar, Shailesh},
  journal={arXiv preprint arXiv:2409.17765},
  year={2024}
}

@article{israel1966singular,
  title={Singular hypersurfaces and thin shells in general relativity},
  author={Israel, Werner},
  journal={Il Nuovo Cimento B (1965-1970)},
  volume={44},
  number={1},
  pages={1--14},
  year={1966},
  publisher={Springer}
}

@article{tan2025black,
  title={Black hole tunneling in loop quantum gravity},
  author={Tan, Hongwei and Guo, Rongzhen and Zhang, Jingyi},
  journal={Chinese Physics C},
  volume={49},
  number={5},
  pages={055106},
  year={2025},
  publisher={IOP Publishing}
}

@article{tan2025massive,
  title={Massive particles tunneling from quantum Oppenheimer-Snyder black holes and black hole entropy},
  author={Tan, Hongwei and Xiao, Kui},
  journal={Physics Letters B},
  pages={139741},
  year={2025},
  publisher={Elsevier}
}

@article{harmark2010greybody,
  title={Greybody factors for d-dimensional black holes},
  author={Harmark, Troels and Natario, Jose and Schiappa, Ricardo},
  year={2010}
}

@article{belfaqih2025hawking,
  title={Hawking evaporation and the fate of black holes in loop quantum gravity},
  author={Belfaqih, Idrus Husin and Bojowald, Martin and Brahma, Suddhasattwa and Duque, Erick I},
  journal={Physical review letters},
  volume={135},
  number={16},
  pages={161501},
  year={2025},
  publisher={APS}
}

@article{parikh2025quantum,
  title={Quantum uncertainty in the area of a black hole},
  author={Parikh, Maulik and Pereira, Jude},
  journal={Journal of High Energy Physics},
  volume={2025},
  number={9},
  pages={1--27},
  year={2025},
  publisher={Springer}
}

@article{iyer1987black,
  title={Black-hole normal modes: A WKB approach. I. Foundations and application of a higher-order WKB analysis of potential-barrier scattering},
  author={Iyer, Sai and Will, Clifford M},
  journal={Physical Review D},
  volume={35},
  number={12},
  pages={3621},
  year={1987},
  publisher={APS}
}

@article{umemoto2018entanglement,
  title={Entanglement of purification through holographic duality},
  author={Umemoto, Koji and Takayanagi, Tadashi},
  journal={Nature Physics},
  volume={14},
  number={6},
  pages={573--577},
  year={2018},
  publisher={Nature Publishing Group UK London}
}

@article{engelhardt2022canonical,
  title={Canonical purification of evaporating black holes},
  author={Engelhardt, Netta and Folkestad, {\AA}smund},
  journal={Physical Review D},
  volume={105},
  number={8},
  pages={086010},
  year={2022},
  publisher={APS}
}

@article{bekenstein1973black,
  title={Black holes and entropy},
  author={Bekenstein, Jacob D},
  journal={Physical Review D},
  volume={7},
  number={8},
  pages={2333},
  year={1973},
  publisher={APS}
}

@article{mathur2009information,
  title={The information paradox: a pedagogical introduction},
  author={Mathur, Samir D},
  journal={arXiv preprint arXiv:0909.1038},
  year={2009}
}

@article{raju2022lessons,
  title={Lessons from the information paradox},
  author={Raju, Suvrat},
  journal={Physics Reports},
  volume={943},
  pages={1--80},
  year={2022},
  publisher={Elsevier}
}

@article{witten2025introduction,
  title={Introduction to black hole thermodynamics},
  author={Witten, Edward},
  journal={The European Physical Journal Plus},
  volume={140},
  number={5},
  pages={430},
  year={2025},
  publisher={Springer}
}

@article{susskind2012singularities,
  title={Singularities, firewalls, and complementarity},
  author={Susskind, Leonard},
  journal={arXiv preprint arXiv:1208.3445},
  year={2012}
}

@article{zhang2025information,
  title={The information loss problem and Hawking radiation as tunneling},
  author={Zhang, Baocheng and Corda, Christian and Cai, Qingyu},
  journal={Entropy},
  volume={27},
  number={2},
  pages={167},
  year={2025},
  publisher={MDPI}
}

@article{han2023geometry2,
  title={Geometry of the black-to-white hole transition within a single asymptotic region},
  author={Han, Muxin and Rovelli, Carlo and Soltani, Farshid},
  journal={Physical Review D},
  volume={107},
  number={6},
  pages={064011},
  year={2023},
  publisher={APS}
}
\end{document}